\documentclass[aps,prb,twocolumn,groupedaddress]{revtex4} %% two column
\usepackage{amsmath}
\usepackage{graphicx}
\begin{document}

%%\twocolumn[ %% activate for two-column option

\title{Effective optical constants in stratified 
metal-dielectric metameterial}

\author{Masanobu Iwanaga}
\affiliation{Department of Physics, Graduate School of Science, 
Tohoku University, Sendai 980-8578, Japan}

\date{\today}

\begin{abstract}
We present effective optical constants of 
stratified metal-dielectric metameterial. The effective 
constants are determined by two complex reflectivity method (TCRM). 
TCRM reveals full components of effective permittivity and permeability 
tensors and indicates the remarkable anisotropy of metallic and dielectric 
components below effective plasma frequency. On the other hand, 
above the plasma frequency, one of the effective refractive 
indexes takes a positive value less than unity and 
is associated with small loss. The photonic states are 
confirmed by the distribution of electromagnetic fields. 
\end{abstract} 
\maketitle
%%\ocis{000.4430, 160.4760, 260.2130, 260.5740.}

%%] %% activate for two-column option

%\noindent \underline{\it Introduction} 

Photonic metamaterials attract great 
interest as a new type of materials including magnetic resonance 
at optical frequency. A famous example exploiting magnetic resonance 
is negative refraction.\cite{Pendry04}

Evaluation of effective optical constants at optical frequency 
has been performed for metamaterials of thin layers.\cite{Smith02} 
In analyzing bulk metamaterials, reflective polarimetry is only way to obtain 
effective optical constants. It is necessary in the analysis to 
satisfy equation of dispersion. 
In this Letter we present a way of analysis, two complex reflectivity method 
(TCRM), to determine full components of effective 
tensors of permittivity $\varepsilon$ and permeability $\mu$. 
Moreover, the photonic states implied by the effective optical 
constants are examined. 

\begin{figure}[b]
\centerline{\includegraphics[width=5.5cm,clip]{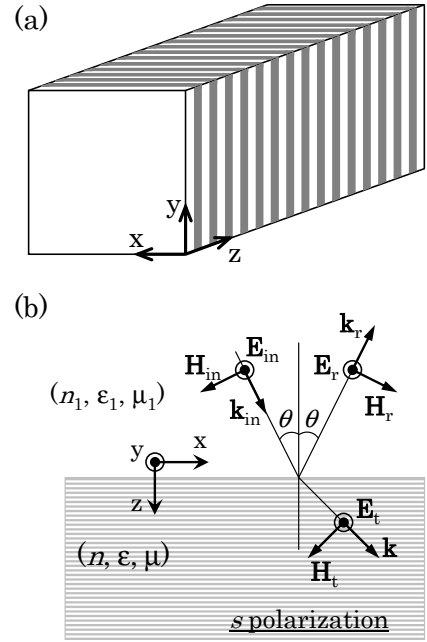}}
\caption{(a) Schematic drawing of SMDM and coordinate 
configuration. Gray indicates metal (silver) layers and write is dielectric 
(MgF$_2$) layers. (b) Optical configuration of TCRM. Incident light is 
$s$ polarization, that is, $\textbf{E}_{\rm in}||y$.}
\label{fig1}
\end{figure}

Figure \ref{fig1} shows stratified metal-dielectric metamaterial (SMDM) 
and coordinate system. 
The composite material is obviously uniaxial and the $z$ axis is set parallel 
to the optical axis. We set here metal to be Ag and dielectrics 
MgF$_2$. The pair of constituents is selected to reduce loss in materials and 
the interfaces.\cite{Dolling}

%\noindent \underline{\it Method}

We assume that effective tensors $\varepsilon$ and $\mu$ 
are diagonal and describe local response in media. Principal axes 
of both tensors are set to be $x$, $y$, and $z$ axes in Fig.\ \ref{fig1}(a). 
In the configuration shown in Fig \ref{fig1}(b), 
incident light is $s$ polarization (that is, $\textbf{E}_{\rm in}||y$), and 
the following relation is derived from Maxwell boundary conditions 
concerning bulk material in vacuum ($n_1 = 1$, $\varepsilon_1 = 1$, and 
$\mu_1 = 1$): 
\begin{equation}
\frac{\hat{k}_z(\theta)}{\mu_x} = \frac{n_1\cos\theta}{\mu_1}\cdot%
\frac{1- r_s(\theta)}{1 + r_s(\theta)}. \label{kz_s}
\end{equation}
We set $\textbf{E}(r,t) \propto \exp(i\textbf{k}\cdot\textbf{r} - i\omega t)$ 
and $\hat{\textbf{k}} = \textbf{k}/k_0$ where $k_0$ is wavenumber of light in 
vacuum. The component $\hat{k}_z$ represents the refraction; in particular, 
$\hat{k}_z(0)$ is refractive index along $z$ axis. 
Complex reflectivity $r_s$ was obtained by the 
numerical calculation based on scattering-matrix method\cite{Tikh} improved 
in numerical convergence.\cite{Li1} Optical constants of silver were taken 
from literature,\cite{Johnson} and the refractive index of MgF$_2$ was set 
1.38. 

Equation of dispersion under $s$ polarization is 
\begin{equation}
\varepsilon_y = \frac{\hat{k}_z(\theta)^2}{\mu_x} + %
\frac{\hat{k}_x(\theta)^2}{\mu_z} \label{s-relation}
\end{equation}
where $\hat{k}_x(\theta) = n_1 \sin\theta$. 

After substituting Eq.\ (\ref{kz_s}) for Eq.\ (\ref{s-relation}), 
two different angles 
enable to evaluate $\varepsilon_y / \mu_x$ and $\mu_x \mu_z$ uniquely. 
The products of $\varepsilon_z / \mu_y$ and $\mu_y \mu_x$ are obtained 
by permutating configuration $(x,y,z) \to (y,z,x)$. In uniaxial media of 
$\varepsilon_x = \varepsilon_y$ and $\mu_x = \mu_y$, 
only the two configurations are enough to 
determine all the values of tensors of $\varepsilon$ and $\mu$ 
except for the sign. We call this procedure TCRM. 
This method is generalization of two reflectance method (TRM),%
\cite{Azzam79,Azzam80} which is valid for materials of $\mu = 1$. 
In TRM, conformal mapping between reflectance at two incident angles plays 
a crucial role and retrieves the phase of complex reflectivity. 
On the other hand, 
if it is assumed that $\mu \neq 1$, we need two complex reflectivity to 
evaluate $\varepsilon$ and $\mu$ tensors. Thus, though TCRM stems from TRM, 
the formalism is quite different. 

\begin{figure}[t]
\centerline{\includegraphics[width=7.8cm,clip]{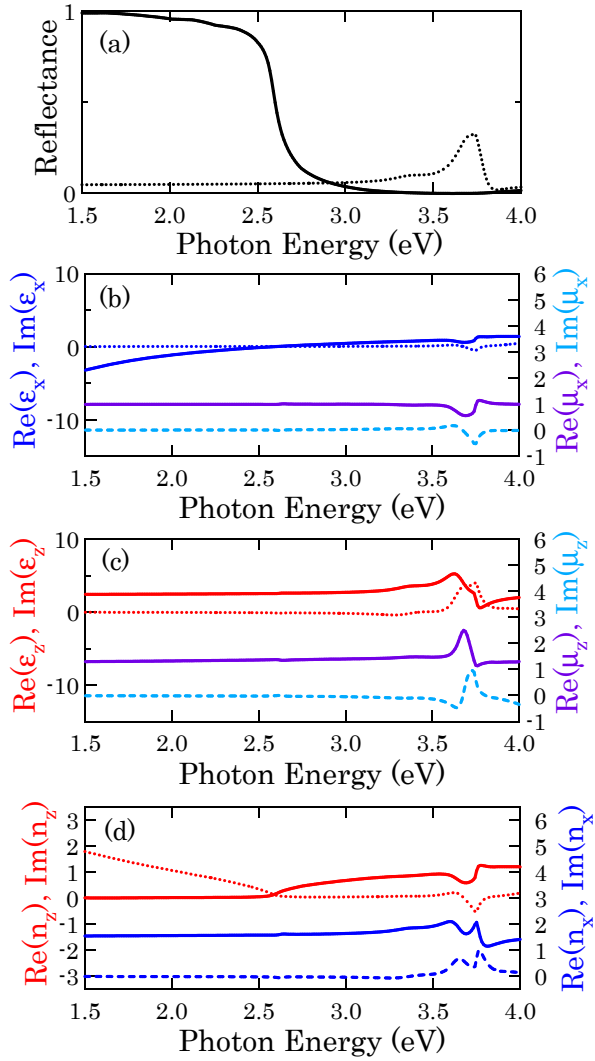}}
\caption{(a) Reflectance spectra. Solid line: under normal incidence on 
$xy$ plane. Dotted line: under nornal incidence with 
$\textbf{E}_{\rm in}||z$ on $yz$ plane. 
(b) The $x$ component of effective $\varepsilon$ and $\mu$. 
Upper solid line denotes Re($\varepsilon_x$) and dotted line does 
Im($\varepsilon_x$). Lower solid line shows Re($\mu_x$) and dashed line does 
Im($\mu_x$).
(c) The $z$ component of effective $\varepsilon$ and $\mu$. 
The notations are similar to (c): replace $x$ in (c) with $z$.
(d) Refractive indexes. Upper solid line depicts Re($n_z$) and dotted line 
does Im($n_z$). Lower solid line represents Re($n_x$) and dashed line does 
Im($n_x$).}
\label{fig2}
\end{figure}

It is nontrivial how to determine the sign of $\mu_x$. 
Actually we take the sign to satisfy $\mu_x\approx 1$ at 
off-resonance energy of 1.5 eV and to connect $\mu_x$ at resonance 
without discontinuous jump. Once the sign of $\mu_x$ is identified, other 
components of $\mu_z$, $\varepsilon_x$, and $\varepsilon_z$ are evaluated 
uniquely; besides, the refractive component $\hat{k}_z$ is also evaluated 
uniquely by Eq.\ (\ref{kz_s}). 
As shown in Fig.\ \ref{fig2}, this way 
determines effective tensors $\varepsilon$ and $\mu$ 
in a wide energy range. 

%\noindent \underline{\it Results} 

Figure \ref{fig2}(a) shows reflectance spectra in the two configuration for 
TCRM; solid line indicates reflectance spectra under normal 
incidence on $xy$ plane and dotted line under normal and 
$\textbf{E}_{\rm in}||z$ incidence on $yz$ plane. 
The periodicity of SMDM is 75 nm, and thickness of Ag and MgF$_2$ are 15 
and 60 nm, respectively. Surface layer parallel to $xy$ plane is set to be 
MgF$_2$ and the thickness is set 30 nm. 
Reflectance spectrum shown by the solid 
line was evaluated for thick SMDM of 2000 periods in numerical evaluation, 
and reflectance drawn with dotted line was calculated 
for 50 mm thick SMDM. 
The thickness are enough to ensure that the two reflectance (and $r_s$) 
are numerically same with infinitely thick SMDM. 
Numerical accuracy is tested with changing the number of Fourier harmonics 
employed in the numerical calculation for $r_s$ 
and is estimated within a few percent. 

\begin{figure}[b]
\centerline{\includegraphics[width=7.8cm,clip]{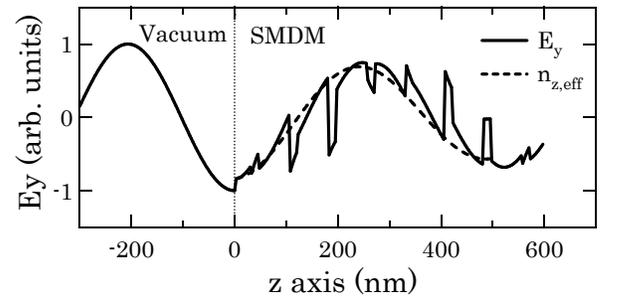}}
\caption{(a) Effect of refractive index $n_z$ at 3.0 eV. 
Solid line represents the electric field $E_y(z)$ of incident light 
($\textbf{E}_{\rm in}||y$, normal incidence on $xy$ plane) in vacuum 
and the refracted component in SMDM. 
Dashed line: the $E_y(z)$ evaluated by using $n_z$; 
Fig.\ \ref{fig2}(d) shows $n_z = 0.67+0.04i$ at 3.0 eV.}
\label{fig3}
\end{figure}

Figures \ref{fig2}(b) and (c) respectively displays $x$- and $z$-components 
of effective $\varepsilon$ and $\mu$. 
In Figs.\ \ref{fig2}(b) and (c), upper solid lines denote Re($\varepsilon_i$), 
and lower solid lines stand for Re($\mu_i$) ($i=x,z$). 
In this TCRM analysis, two angles of 0 and 15 degrees are employed. 
Other pairs of angles derive same results within the numerical errors. 
It is therefore confirmed that 
TCRM determines full components of $\varepsilon$ and $\mu$ 
with satisfying equation of dispersion in SMDM. The component 
$\varepsilon_x$ represents typical permittivity of Drude metal below 3.5 eV. 
The effective plasma frequency $\omega_{\rm p,eff}$ 
is located at $\hbar\omega_{\rm p,eff} = 2.6$ eV. 
On the other hand, $\varepsilon_z$ 
implies that SMDM is dielectrics for the light of $\textbf{E}||z$. 
Below 3.5 eV there exist no prominent magnetic resonance in SMDM, 
while magnetic 
resonance appears at 3.7 eV. The resonance deviates from simple Lorentzian 
dispersion and is associated with resonant behavior of 
permittivity. The 3.7 eV resonance is thus mixed one of electromagnetic (EM) 
components and shows complex response due to anisotropy of SMDM. 
This resonant behavior is discussed later in more details. 

Figure \ref{fig2}(d) presents effective refractive indexes. 
SMDM shows a metallic color similar to gold in seeing from $xy$ plane, 
while it is transparent dielectric material of refractive index 1.8 
in visible range for the light of $\textbf{E}||z$ in $yz$ or $xz$ planes. 
In particular, loss in SMDM is quite suppressed in the wide range of 
2.6-3.5 eV. This is 
a feature of SMDM. Although the permittivity of bulk silver implies large loss 
in the energy range, the composite of silver and MgF$_2$ is nearly free from 
loss. This result suggests that EM fields in metallic layers 
help transmission highly efficient; that is, it suggests that 
there exists EM-field enhancement. 
Indeed, above $\omega_{\rm p,eff}$, several layers of metal and dielectrics 
shows high transmittance.\cite{Scal98} 
Figure \ref{fig2}(d) indicates that high-efficient transmission 
in SMDM is associated with effective refractive index $n_z$ of 
$0 < \textrm{Re}(n_z) < 1$. The effective index thus provides a concise 
description for the photonic state. Next, we examine EM-field 
distribution in SMDM to clarify whether the state is actually realized. 

Figure \ref{fig3} displays the $E_y$ profile along $z$ axis at 3.0 eV. 
Incident light travels along $z$ axis and illuminates on $yz$ plane with 
$\textbf{E}_{\rm in}||y$. Solid line shows incident light in vacuum and 
the refracted component in SMDM. 
In this configuration the effective refractive index is $n_z = 0.67+0.04i$ 
from Fig.\ \ref{fig2}(d). The $E_y$ reproduced by $n_z$ is drawn with dashed 
line. The wavelength in SMDM is obviously 
close to $\lambda/$Re($n_z$) and longer that $\lambda$ 
($\lambda$: wavelength in vacuum). 
Consequently, it is definitely confirmed 
that the effective description for SMDM works well. 
In particular, the effective description is excellent agreement within a 
$\lambda/\{4$Re($n_z$)\} scale from the interface, and suggests that 
TCRM mainly analyzes the EM responses within the $\lambda/\{4$Re($n_z$)\} 
depth. 
The index $n_z$ represents the low-loss photonic state in a wide range 
of 2.6-3.5 eV, where the phase velocity exceeds the velocity of light $c$ 
in vacuum. Figure \ref{fig3} also shows that 
the EM wave in SMDM is not ideal plane wave and is affected by periodic 
structure. It is one of the inevitable properties in mesoscopic metamaterial 
optics. 

\begin{figure}[tb]
\centerline{\includegraphics[width=7.8cm,clip]{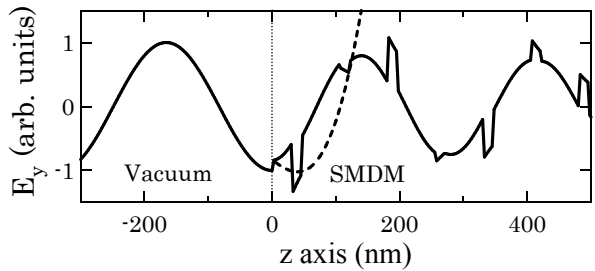}}
\caption{Profile of electric field $E_y$ at 3.74 eV (solid line) 
under normal incidence on $xy$ plane. 
Dashed line is the profile calculated from the effective refractive 
index $n_{\rm z}$ at 3.74 eV in Fig.\ \ref{fig2}(d). 
Weak reflection is not included in vacuum for simplicity.}
\label{fig4}
\end{figure}

At the end of discussion, we refer to the unusual behavior of Im($n_z$) 
at 3.7 eV in Fig.\ \ref{fig2}(d). At the resonance, Im($n_z$) takes negative 
value and seems to imply exponential growing wave. However, 
as shown in Fig.\ \ref{fig4}, the EM distribution at 3.7 eV (solid line) 
does not show any such growing wave, and indicates EM-field 
enhancement in metallic layers. 
The index $n_z$ implies the profile drawn with dashed line in 
Fig.\ \ref{fig4}, and roughly reproduces the EM fields in SMDM 
within a half wavelength scale from the interface. 
Why does such unusual behavior appear? 
In Fig.\ \ref{fig4}, the wavelength in vacuum is $\lambda = 331$ nm and 
the periodicity of SMDM is 75 nm. 
It therefore seems still possible to use the effective 
description; however, the other component is Re($n_x$) = 2.0 
and $\lambda$/Re($n_x$) = 166 nm which is about twice the periodicity. 
The condition in Fig.\ \ref{fig4} is likely close to 
limits of effective description. 
Additionally we note that the limits of effective description were eagerly 
debated in GHz range;\cite{KMSS,DepineEfros} one of the limits 
is at present understood to be 
$\lambda/|{\rm Re}(n_{\rm eff})| \sim$ periodicity.

From numerical results and discussion, we can extract a few lessons in 
TCRM analysis: 
(i) as for wavelength $\lambda$ in vacuum and periodicity $a$ in 
metamaterial, it is at least necessary that $\lambda/|$Re($n_{\rm eff}$)$|$
$ > 2a$, in order to obtain usual effective refractive index, 
Im($n_{\rm eff}$)$ \ge 0$. We state that this condition is satisfied below 
3.5 eV in Fig.\ \ref{fig2} and that both $\varepsilon$ and $\mu$ are 
determined successfully; 
(ii) effective optical constants are determined by the optical response 
mainly within a $\lambda/|4{\rm Re}(n_{\rm eff})|$ scale from the interface. 

%\noindent \underline{\it Conclusions}

In conclusion, we have presented the scheme of TCRM, applied TCRM to SMDM, 
and revealed the full components of $\varepsilon$ and $\mu$. 
Analysis of EM distribution has confirmed that the effective refractive index 
well describes the photonic states in SMDM below 3.5 eV. 
The effective index indicates that the 
low-loss photonic states with phase velocity larger than $c$ exists 
in a wide energy range above $\omega_{\rm p,eff}$.

The author thanks T.\ Ishihara for discussion and comments on manuscript. 
This study is partially supported by Research Foundation for Opto-Science 
and Technology, and by Information Synergy Center, 
Tohoku University in numerical implementation. 

%\bibliography{iwanaga15}

\end{document}